\documentclass[acmsmall,screen]{acmart}

\AtBeginDocument{%
  }

\usepackage{algorithmic}
\usepackage{graphicx}
\usepackage{textcomp}
\usepackage{xcolor}

\usepackage{makecell}
\usepackage{tabularx}
\usepackage[T1]{fontenc}
\usepackage{comment}
\usepackage{xcolor}
\usepackage{multirow}
\usepackage{array}
\usepackage{soul}
\usepackage{threeparttable}
\usepackage{balance}
\usepackage{diagbox}
\usepackage{subfigure}
\usepackage{microtype} 
\usepackage{calc}

\usepackage{enumitem}
\usepackage{url}
\usepackage{xspace}
\usepackage{float}
\usepackage{stfloats}
\usepackage{adjustbox}
\usepackage{amsmath,amssymb,amsfonts}
\usepackage{algorithmic}
\usepackage{graphicx}
\usepackage{textcomp}
\usepackage{xcolor}
\usepackage{multirow}
\usepackage{mdframed}
\usepackage{tcolorbox}
\usepackage{makecell}
\usepackage{threeparttable}
\usepackage{colortbl}
\usepackage{subfigure}
\usepackage{hhline}
\usepackage{pifont}
\usepackage{enumitem}

\newcolumntype{C}{>{\centering\arraybackslash}X}
\usepackage{pgfplots}
\pgfplotsset{compat=1.16}
\usetikzlibrary{pgfplots.statistics}

\newcommand{\colortext}{\textcolor{black}}
\newcommand{\parabf}[1]{\noindent\textbf{#1}}

\definecolor{ggray}{HTML}{eff0f0}
\definecolor{gggray}{HTML}{E8E8E8}
\definecolor{ggggray}{HTML}{BEBEBE}

\newcommand{\claude}{Claude-3-5-Sonnet}
\newcommand{\vdeepseek}{DeepSeek-V3-0324}
\newcommand{\rdeepseek}{DeepSeek-R1}
\newcommand{\qwenthreetwo}{Qwen-2.5-coder-32B}
\newcommand{\qwenfourteen}{Qwen-2.5-coder-14B}

\newcommand{\astmatchscore}{Match$_{ast}$\xspace}
\newcommand{\ourapproach}{K$^{\mathsf{3}}$Trans\xspace}

\newcommand{\ie}{\textit{i.e.,}\xspace}

\definecolor{myyellow}{HTML}{FFF2CC}

\newcounter{finding}
\newcommand{\finding}[1]{\refstepcounter{finding}
 	\vspace{1mm}
	\begin{mdframed}[linecolor=gray!25,roundcorner=12pt,backgroundcolor=myyellow!30,linewidth=3pt,innerleftmargin=2pt, leftmargin=0cm,rightmargin=0cm,topline=false,bottomline=false,rightline = false]
		\textbf{Finding \arabic{finding}:} #1
	\end{mdframed}
	\vspace{1mm}
}

\newcommand{\boxmargin}{1mm}
\tcbuselibrary{most, skins, breakable, theorems}
\newtcolorbox{myboxa}[2][]{
    colback=gray!10!white,
    colframe=black, enhanced,
    attach boxed title to top left={yshift=-2mm,xshift=5mm},
    title=#2,#1
}
\newtcolorbox{myboxb}[2][]{
    boxsep=3pt,
    left = \boxmargin, right = \boxmargin, top = \boxmargin, bottom = \boxmargin,
    title={#2},#1
}
\newtcolorbox{myboxc}{
    colback=gray!15!white,
    arc = 0pt, outer arc = 0pt,
    boxsep=0pt, left = 3pt, right = 0pt, top = 0pt, bottom = 0pt, 
    leftrule=3pt, bottomrule=0pt,toprule=0pt, rightrule=0pt,
    left = \boxmargin, right = \boxmargin, top = \boxmargin, bottom = \boxmargin
}
\newtcolorbox{myboxd}{
    colback=gray!10,
    colframe=black,
    width=\columnwidth,
    arc=1mm, auto outer arc,
    boxrule=0.5pt,
}

\usepackage{tikz}

\begin{document}

\title{\ourapproach: Evolving Triple Knowledge-Augmented LLMs for Code Translation in Repository Context}

\author{Guangsheng Ou}
\affiliation{%
  \institution{Sun Yat-sen University}
  \country{China}}
\email{ougsh3@mail2.sysu.edu.cn}

\author{Mingwei Liu*}
\affiliation{%
  \institution{Sun Yat-sen University}
  \country{China}}
\email{liumw26@mail.sysu.edu.cn}

\author{Yuxuan Chen}
\affiliation{%
  \institution{Sun Yat-sen University}
  \country{China}}

\author{Xueying Du}
\affiliation{%
  \institution{Fudan University}
  \country{China}}

\author{Shengbo Wang}
\affiliation{%
  \institution{Sun Yat-sen University}
  \country{China}}

\author{Zekai Zhang}
\affiliation{%
  \institution{Sun Yat-sen University}
  \country{China}}

\author{Xin Peng}
\affiliation{%
  \institution{Fudan University}
  \country{China}}
  
\author{Zibin Zheng}
\affiliation{%
  \institution{Sun Yat-sen University}
  \country{China}}

\begin{abstract}
Large language models (LLMs) have behaved well in function-level code translation without repository-level context. 
However, the performance of LLMs in repository-level context code translation remains suboptimal due to complex dependencies and context, hindering their adoption in industrial settings.
In this work, we propose a novel LLM-based code translation technique \ourapproach, which leverages triple knowledge augmentation to enhance LLM's translation quality under repository context in real-world software development.
First, \ourapproach constructs an evolving translation knowledge base by extracting relevant information from target-language codebases, the repository being translated, and prior translation results.
Second, for each function to be translated, \ourapproach retrieves relevant triple knowledge, including target-language code samples, dependency usage examples, and successful translation function pairs, serving as references to enhance LLM for translation.
Third, \ourapproach constructs a knowledge-augmented translation prompt using the retrieved triple knowledge and employs LLMs to generate the translated code while preserving repository context. It further leverages LLMs for self-debugging, enhancing translation correctness.
\colortext{Lastly, \ourapproach continuously evolves the translation knowledge base by extracting new dependency usage examples and successful translation function pairs from prior translation history or from components that have been refactored, or updated manually by developers and extracting target language code samples by automatically detecting and downloading newly added target-language project.}

The experiments show that \ourapproach substantially outperforms the baseline adapted from previous work, achieving relative improvements of up to 135.9\% on Pass@1 and 32.8\% on CodeBLEU among studied LLMs. Furthermore, the code generated by \ourapproach is of higher quality, as indicated by the higher DSR@1 and Repairable Ratio, which suggests a greater proportion of fixable code.
It should be noted that the results also demonstrate that each knowledge significantly contributes to \ourapproach’s effectiveness in handling repository-level context code translation, with dependency usage examples making the most notable contribution. 
Moreover, as the self-evolution process progresses, the knowledge base continuously enhances the LLM's performance across various aspects of the repository-level code translation.

\end{abstract}



\maketitle

\vspace{-5pt}
\section{Introduction}

\begin{figure*}
    \centering
    \includegraphics[width=1\linewidth]{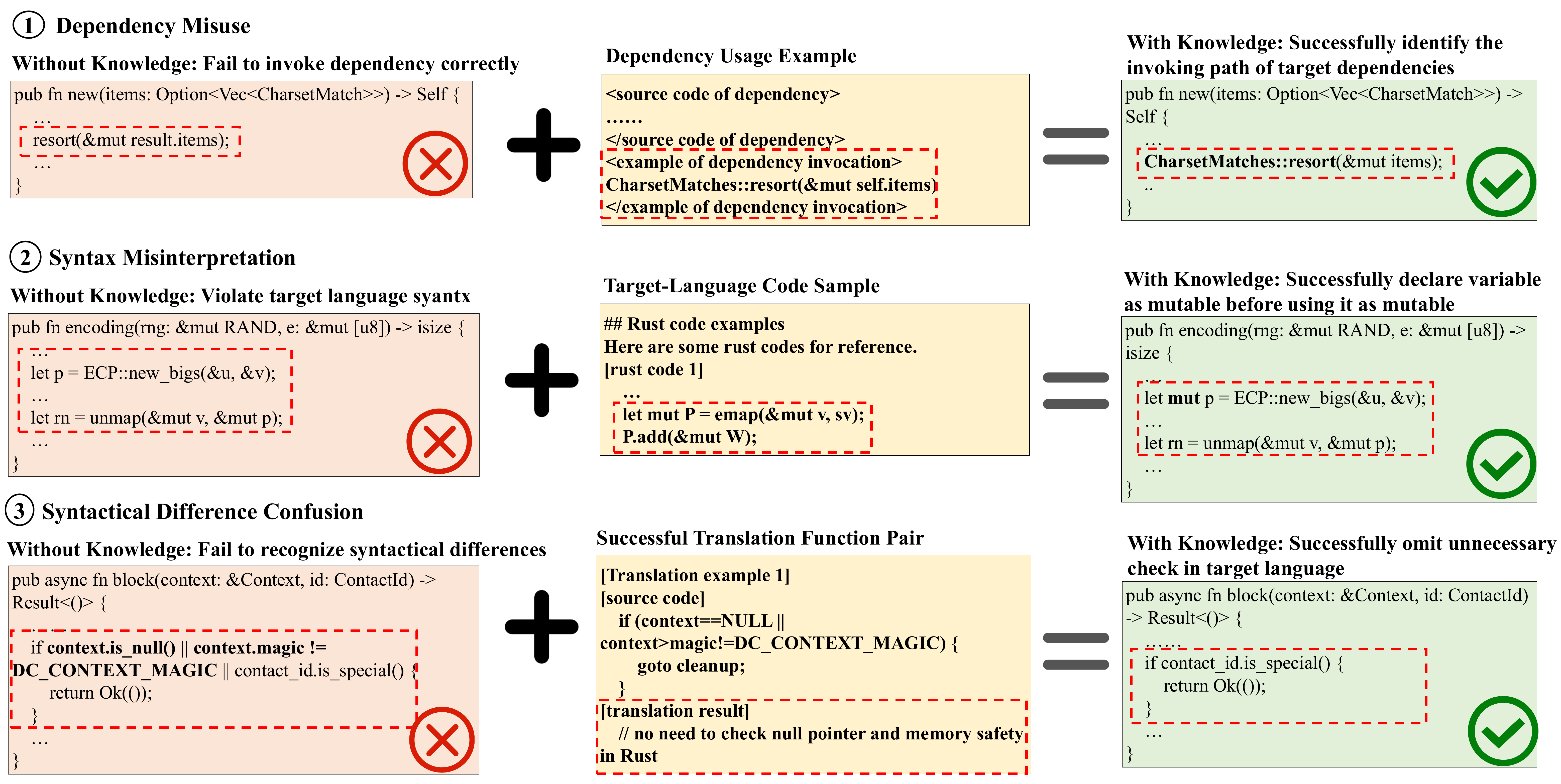}
    \vspace{-15pt}
    \caption{Motivational Examples of LLMs' Performance With vs. Without Knowledge in Addressing Three Key Challenges: Dependency Misuse, Syntax Misinterpretation, and Syntactical Difference Confusion}
    \label{fig:motivation_example}
    \vspace{-15pt}
\end{figure*}

Code translation plays a critical role in modern software development, as manual translation is both time-consuming and error-prone, necessitating robust automated tools~\cite{jiao2023evaluation}. Recent advances in large language models (LLMs) have demonstrated remarkable performance in various natural language processing tasks, including code translation~\cite{pan2024lost, yang2024exploring, tao2024unraveling}. Conventional benchmarks, such as CodeTransOcean~\cite{yan2023codetransocean}, have shown that LLMs perform well on function-level translation tasks. 

However, these benchmarks largely ignore the challenges posed by repository-level context, where \textbf{real-world code translation must consider complex dependencies, project-specific coding styles, and evolving repository structures}~\cite{ou2024repository, zhang2025skeletonguidedtranslationbenchmarkingframeworkcode, wang2024repotransbenchrealworldbenchmarkrepositorylevel}. 
\colortext{Furthermore, in real-world projects, translation rarely rewrites an entire repository at once. Developers typically incrementally translate new or modified modules from the source to the target language~\cite{zhang2023multilingual, incremental_translation_1, migrate_to_new_language1, SmaCC_2009}. Newly translated code must integrate with existing modules, including components migrated earlier or manually refined, to preserve functional correctness and consistency. Incremental workflows also arise when the source project evolves after partial migration, requiring new features to be ported into the target repository. \textbf{Such repository-level migration demands careful handling of non one-to-one mapping repository-level context such as different dependencies, distinct function signatures, and architectural divergence.}}
As a result, the performance of LLMs \colortext{in incremental translation scenario with evolving repository-context} remains below expectations, limiting their practical adoption in industrial applications. Ou et al.~\cite{ou2024repository} demonstrates that even the best model experiences a 30.8\% performance drop (Pass@1 from 74.3\% to 43.5\%) when handling repository-level context compared to previous benchmarks without such context.

Figure~\ref{fig:motivation_example} illustrates three translation examples drawn from the evaluation results of LLMs on the \textbf{RustRepoTrans} benchmark that highlight common translation \colortext{challenges}.
(1) \textbf{Dependency Misuse}: An LLM fails to correctly apply dependency usage, leading to errors. Injecting a relevant dependency usage example from the repository into the translation prompt resolves this issue. (2) \textbf{Syntax Misinterpretation}: LLM produces code that violates target language syntax (e.g., error ``cannot borrow `p' as mutable, as it is not declared as mutable''). Providing an example from an open-source project with similar syntax characteristics guides the LLM toward correct syntax. (3) \textbf{Syntactical Difference Confusion}: An LLM fails to recognize syntactical differences between the source language and the target language, resulting in unnecessary checks, such as redundant null pointer or memory safety validations, which are necessary in the source language, but do not align with the target language’s paradigms. A reference to successful translation function pairs, where such checks are omitted, enables the LLM to avoid these pitfalls.
Existing work~\cite{bhattarai2024enhancingcodetranslationlanguage} using RAG to enhance LLM code translation has only provided code translation pairs as a single type of knowledge. It does not consider repository context information in real-world software development scenarios or the fact that the target language in practical translation tasks is often an emerging, low-resource language~\cite{ou2024repository}.
\colortext{Furthermore, the retrieval scope of existing methods is limited to a static knowledge base, which does not evolve dynamically as the translation process advances.}
Consequently, previous approaches overlook the importance of dependency usage knowledge and target language syntax knowledge, resulting in poor performance in real translation scenarios. 
\colortext{Additionally, the static nature of the knowledge base restricts the range of tasks that RAG can support, making it ineffective in handling continuously evolving repository-level context.}

Motivated by these challenges, this work proposes \ourapproach, a novel LLM-based code translation technique that leverages \textbf{\colortext{evolving} triple knowledge augmentation} to enhance translation quality under repository context. Specifically, \ourapproach integrates: (1) \textbf{Dependency Usage Examples} extracted from the repository being translated, offering insights into proper dependency invocation; (2) \textbf{Target-language Code Samples} drawn from existing projects, which provide real-world syntax and style references;  (3) \textbf{Successful Translation Function Pairs} from translation history, serving as concrete examples for correct code transformation.

By building a continuously updating dynamic knowledge base, \ourapproach adapts to evolving repository contexts and improves the LLM’s ability to generate accurate, context-aware code. Our experiments, conducted on the newly repository-level context code translation benchmark \textbf{RustRepoTrans}, demonstrate that incorporating triple knowledge not only mitigates common translation errors but also significantly enhances overall translation performance. 
First, we compare \ourapproach with LLM-based baselines: PE and Basic RAG adapted from~\cite{bhattarai2024enhancingcodetranslationlanguage}, \colortext{and the Rule-based baseline: C2Rust}~\cite{c2rust}. The results show that \ourapproach outperforms all the baselines across all match-based and execution-based metrics. Specifically, compared to Basic RAG, \ourapproach achieves relative improvements of up to 135.9\% in Pass@1 and 32.8\% in CodeBLEU among the studied LLMs. Furthermore, even the prompt design in \ourapproach alone slightly outperforms the baseline adapted from prior work. \colortext{Moreover, C2Rust only successfully translated a single case out of 145 tasks, demonstrating that the rule-based baseline is incapable of handling non one-to-one mapping repository-level context, which is common in real-world scenarios.}
Second, we explore the impact of different types of knowledge in the knowledge base on \ourapproach by progressively reducing the provided knowledge categories. The results demonstrate that each knowledge significantly contributes to \ourapproach’s effectiveness in handling repository-level context code translation, with dependency usage examples making the most notable contribution. 
Third, we evaluate the impact of the self-evolution process on \ourapproach and find that as the self-evolution process progresses, the knowledge base continuously enhances the LLM’s performance across various aspects of the repository-level code translation.
Lastly, we manually analyze the successful examples of \ourapproach that the baselines fail, and the failure examples of \ourapproach to fully understand the advantages and limitations of \ourapproach.

In summary, our contributions are as follows:
\vspace{-3pt}

\begin{itemize}  
    \item \textbf{Triple Knowledge Base Construction}: We propose a self-evolving method to construct three complementary knowledge bases—dependency usage examples, target-language code samples, and successful translation function pairs—to enhance repository-level code translation.  
    \item \textbf{Triple Knowledge-Augmented Translation}: We introduce a novel LLM-based translation approach that integrates triple knowledge to improve translation quality while preserving repository context.  
    \item \textbf{Extensive Evaluation}: We conduct comprehensive experiments on the RustRepoTrans benchmark, demonstrating significant improvements among all studied LLMs over existing methods in accuracy and robustness.  
\end{itemize}

\vspace{-15pt}
\section{Approach}
\label{sec:app}

\begin{figure*}[]
    \centering
    \includegraphics[width=1\linewidth]{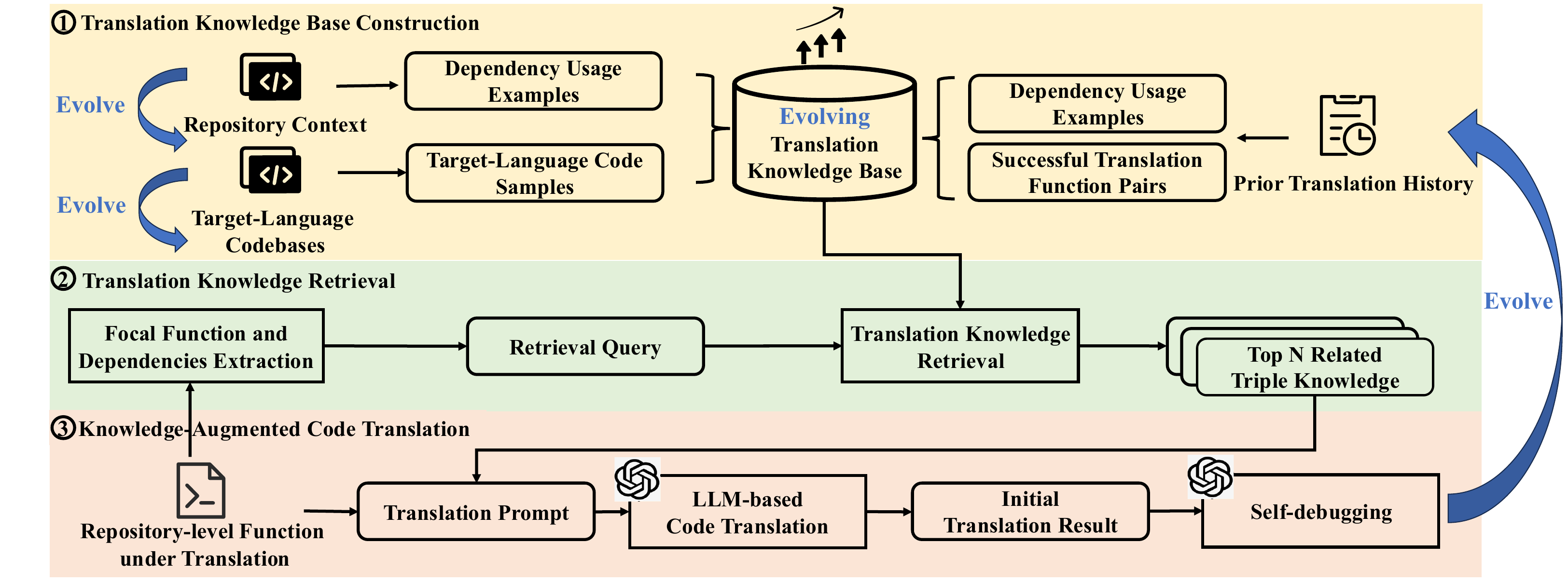}
    \vspace{-20pt}
    \caption{Overview of \ourapproach}
    \label{fig:overview_of_our_apporoach}
    \vspace{-25pt}
\end{figure*}

\subsection{Overview}
To improve code translation within a repository context, this work introduces \ourapproach, an LLM-based technique that enhances translation through triple knowledge augmentation from three key sources:
(1) dependency usage examples from the repository being translated,
(2) target-language code samples from existing projects, and
(3) successful translation function pairs from code translation history.
Figure~\ref{fig:overview_of_our_apporoach} presents the framework of \ourapproach. Given a repository-level context function to translate, \ourapproach leverages triple knowledge to generate target-language code. The framework consists of three phases: Phase-1 (Translation Knowledge Base Construction) is performed offline, whereas Phase-2 (Translation Knowledge Retrieval) and Phase-3 (Knowledge-Augmented Code Translation) operate online.

\begin{itemize}[left=0em] 

\item \textbf{Phase-1: Translation Knowledge Base Construction (Section \ref{sec:phase1})}. \ourapproach constructs a translation knowledge base by extracting relevant information from target-language codebases, the repository being translated, and prior translation results.

\item \textbf{Phase-2: Translation Knowledge Retrieval (Section \ref{sec:phase2})}. For each function to be translated, \ourapproach retrieves relevant triple knowledge, including dependency usage examples, target-language code samples, and successful translation function pairs, serving as references for translation.

\item \textbf{Phase-3: Knowledge-Augmented Code Translation (Section \ref{sec:phase3})}. \ourapproach constructs a knowledge-augmented translation prompt using the retrieved triple knowledge from Phase-2 and employs LLMs to generate the translated code while preserving repository context. It further leverages LLMs for self-debugging, enhancing translation correctness.
\end{itemize}

\ourapproach operates as a \textbf{self-evolving} framework, continuously improving translation quality as target-language codebases, repository structures, and translation history evolve over time. By dynamically incorporating up-to-date translation knowledge, \ourapproach enhances the adaptability and accuracy of translation.

\vspace{-5pt}
\subsection{Translation Knowledge Base Construction}
\label{sec:phase1}

In \ourapproach, the translation knowledge base consists of three parts: dependency usage examples from the repository being translated (in Section \ref{sec:dependency_calling_examples}), target-language code samples from existing projects (in Section \ref{sec:target_language_code_examples}), and translation pair examples from previous successful translation (in Section~\ref{sec:successful_translation_pairs}), which are aimed to enhance LLM's ability to correctly recognize and invoke dependencies, understanding of target language's syntax, and capability to identify syntax differences. 

\subsubsection{Dependency Usage Examples Extraction}
\label{sec:dependency_calling_examples}

Since the same dependency is often called multiple times within the same project, the way a dependency is called elsewhere in the project can serve as a useful reference for its usage in the target function. Specifically, when the scope is the same, the same dependency will have the same call path and identical types of input parameters. When the scope differs, the call path will be similar and the input parameters will still be of the same type. 
Therefore, for each dependency involved in a translation task, we extract the corresponding dependency call statements from the code in the current project that shares the same scope as the target function. If multiple call statements are found, only the first occurrence is retained as the example.
Specifically, by utilizing tree-sitter~\cite{tree-sitter}, we extract function dependencies by identifying \textit{call\_expression} nodes to retrieve all function invocation statements. These statements are then filtered based on the names of the target function dependency, obtaining the required dependency invocation statements.
For variable dependencies, we extract code execution statements and filter them according to the names of the target variable dependency, thereby obtaining the corresponding variable dependency invocation statements.
Ultimately, for each dependency, we constructed the corresponding dependency usage examples with \texttt{[source code of dependency]} and \texttt{[usage example]} (if no corresponding invocation statement was retrieved, \texttt{[usage example]} is set to empty). These extracted statements form the dependency usage examples, enhancing the LLM's ability to correctly identify and invoke dependencies.

This component is self-evolving by continuously performing the dependency extraction operations mentioned before \colortext{to extract dependency usage examples from successfully translated code or from components that have been refactored, or updated manually by developers that has been subsequently integrated into the project's repository context during translation process.}

\vspace{-6pt}
\subsubsection{Target-language Code Samples Extraction}
\label{sec:target_language_code_examples}
Existing work has shown that providing the model with code snippets similar to the target code as a reference improves the quality of generated code in code generation tasks~\cite{fan2024surveyragmeetingllms}. 
Since code implementing the same functionality often uses similar function names and variable names, leading to higher textual similarity~\cite{zhong2010mining}, retrieving the target-language code with the highest textual similarity to the function under translation allows us to effectively identify code samples from similar-domain projects that implement comparable functionality.
These code samples could help improve translation as well.
To simulate the most comprehensive reference code for the target language currently available, we obtained open source projects on GitHub that were created before March 31, 2022, from the The Stack dataset~\cite{Kocetkov2022TheStack} to directly assess function-level code samples. Then, using the official GitHub API, we crawled open source projects created after March 31, 2022, based on the condition \textit{`language: \$target\_language created:2022-04-01..2025-02-01'}, and then used tree-sitter~\cite{tree-sitter} to extract function from source file. 
To prevent potential data leakage in the knowledge base, we remove all projects with the same name as those involved in evaluation benchmark~\cite{ou2024repository}, regardless of whether their creators are the same, considering the possibility of project forks.
These two parts together form the General Projects of the knowledge base, containing a total of 113,882,369 target language code examples from 671,240 projects, which are used to enhance the code generated by the LLM to better conform to the syntax requirements of the target language.

This component is self-evolving by automatically detecting and downloading newly added target-language projects from GitHub with pre-configured update interval.

\vspace{-5pt}
\subsubsection{Successful Translation Function Pair Examples Collection}
\label{sec:successful_translation_pairs}
Providing the model with information of the same category as the target task could maximize its performance on the current task. However, due to the scarcity of existing repository-level context code translation datasets~\cite{ou2024repository}, it is difficult to obtain a large number of high-quality function-level equivalence pairs to construct successful translation function pairs base, for repository-level context. Therefore, we collect successful translation function pairs during the translation process and add them to the knowledge base, constructing the successful translation function pairs knowledge base, which are used to enhance model's ability to recognize syntactic differences when performing code translation tasks.

This component is self-evolving by constantly extracting successful translation function pairs, which has been evaluated by test cases, from prior translation history.
\vspace{-5pt}
\subsection{Translation Knowledge Retrieval}
\label{sec:phase2}
\begin{figure*}
    \centering
    \includegraphics[width=1\linewidth]{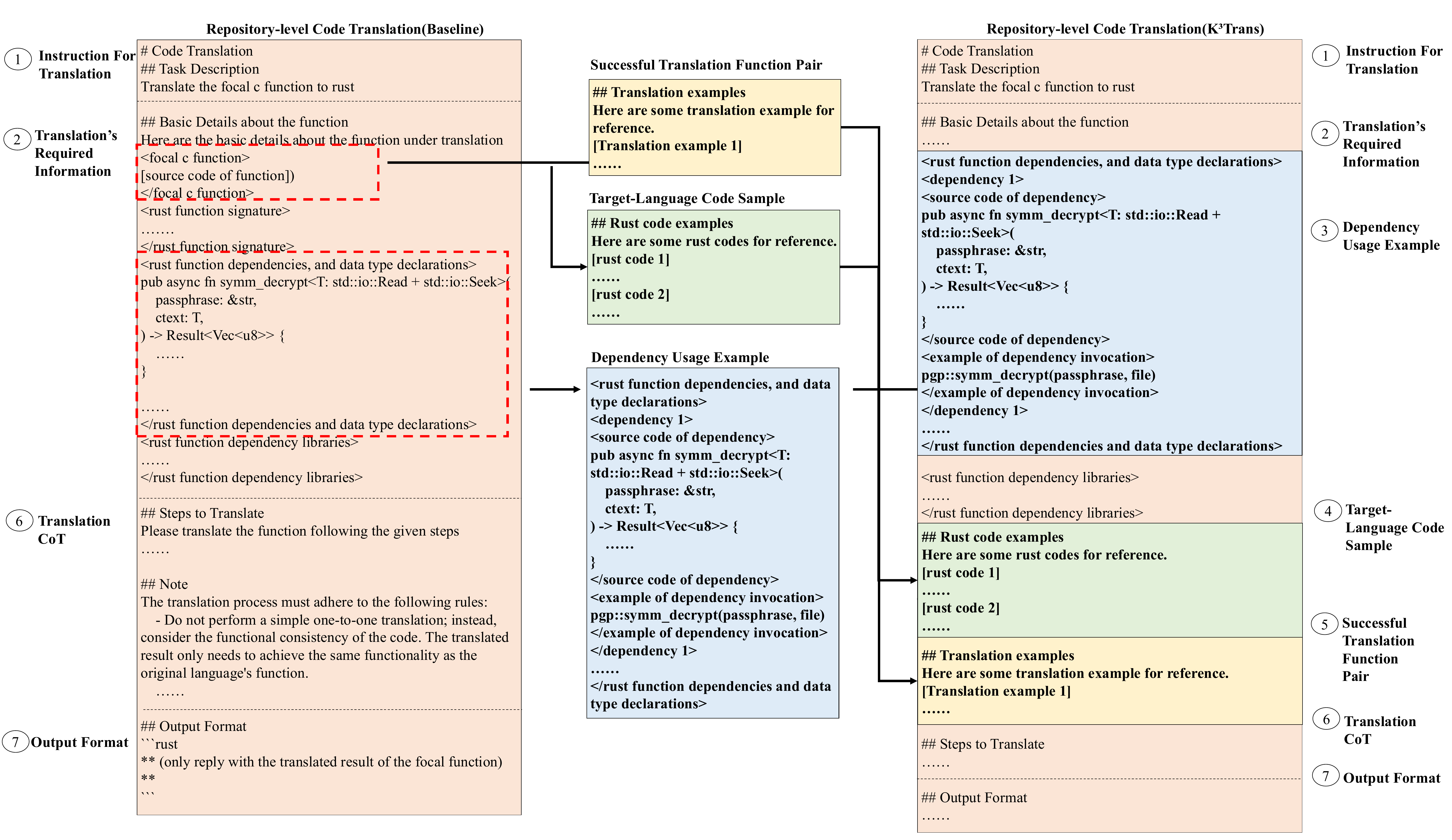}
    \vspace{-18pt}
    \caption{Overview of Prompt Construction (using C to Rust as example)}
    \label{fig:prompt_example}
    \vspace{-13pt}
\end{figure*}

The input of the repository-level context code translation task consists of three parts: the source code of the function under translation, the function signature of target function, and the target function dependencies.
For a given repository-level context code translation task, \ourapproach retrieves relevant translation knowledge from the constructed translation knowledge to construct translation prompt (as shown in Figure~\ref{fig:prompt_example}) base in a three-step retrieval process: focal function and dependencies extraction, candidate knowledge retrieval, candidate knowledge re-ranking.

\vspace{-3pt}
\subsubsection{Focal Function and Dependencies Extraction}
For each repository-level context code translation task, \ourapproach follows the segmentation method described in the evaluated dataset to extract different parts of the translation task. With pattern matching, it separately extracts the source code of the function under translation and its corresponding dependencies.
The source code of the function under translation is used to retrieve target-language code examples and successful translation function pairs (details in Section~\ref{sec:candidate_knowledge_retrieval}). Meanwhile, the dependencies are matched against the \texttt{[source code of dependency]} in the Dependency Usage Examples and the corresponding \texttt{[usage example]} is extracted as the dependency usage example for the dependency.

\vspace{-3pt}
\subsubsection{Candidate Knowledge Retrieval}
\label{sec:candidate_knowledge_retrieval}
For each focal function to be translated, \ourapproach adopts BM25~\cite{robertson1994some}, a method widely used in search engines due to its efficiency and effectiveness~\cite{achiam2023gpt}, based on Elasticsearch~\cite{elasticsearch}, an open source high-performance, RESTful search and analytics engine, to retrieve the top n, where n = 100 for target-language code sample and n = 10 for successful translation function pairs in our experiments, knowledge items for each query function. 
In the field of information retrieval, BM25 calculates the similarity score between a query \( q \) and the documentation \( d \) for retrieval. BM25 computes the similarity score according to the following Equation \ref{equation:BM25}, where \( f(w_i, q) \) is the word \( w_i \)'s term frequency in query \( q \), and \( IDF(w_i) \) is the inverse document frequency of word \( w_i \). The hyperparameters \( k \) (where \( k = 1.2 \) and \( b = 0.75 \)) are employed to normalize term frequencies and control the influence of document length. Prior to calculating the BM25 similarity, both the query and the retrieval documentation undergo essential preprocessing procedures, including tokenization, lemmatization, and stop word removal~\cite{ccaugatayli2015effect}.

\vspace{-5pt}
{\footnotesize
\begin{equation}
\text{Sim}_{BM25}(q, d) = \sum_{i=1}^{n} \frac{IDF(w_i) \times f(w_i, q) \times  (k+1)}{f(w_i, q) + k \left( 1-b+b \times \frac{|q|}{avgdl} \right)} )
\label{equation:BM25}
\end{equation}
}

\subsubsection{Candidate knowledge Re-ranking}
Although BM25 performs excellently in text-based similarity calculations, it struggles to effectively identify noise, such as comments, and other critical information within code, such as the Abstract Syntax Tree (AST). Therefore, \ourapproach employs UniXcoder~\cite{guo2022unixcoder}, a unified cross-modal pre-trained model for programming languages, to re-rank the retrieved knowledge items by calculating cosine similarity. The top n, where n = 2 for target-language code sample (will be discussed further in Section~\ref{sec:rq3}) and n = 1 for successful translation function pairs in our experiments candidate knowledge items with the highest UniXcoder scores are selected as the final knowledge items to be provided to the LLM-based code translation.

\subsection{Knowledge-Augmented Code Translation}
\label{sec:phase3}
We first utilize LLMs to perform translation based on retrieved triple translation knowledge, obtaining initial results, and then apply LLM-based code repair to fix any identified issues.

\parabf{LLM-based code translation.} Based on the retrieved translation knowledge items, \ourapproach leverages LLMs to translate code from the source language to the target language. The prompt design (as shown in Figure~\ref{fig:prompt_example}) adopts an LLM-friendly format: markdown~\cite{prompt_engineering}, and we also design the Chain-of-Thought strategy~\cite{wei2023chainofthoughtpromptingelicitsreasoning}, guiding the LLM to translate the function step by step. Firstly, the LLM is asked to confirm the functionality to be implemented by the current function. Second, to understand the differences between the source and target programming languages, such as the differences of dependencies, syntax and available local variables in detail, we apply the recitation technique, asking the LLM to enumerate all used dependencies and local variables as well as distinguish the syntax differences. Finally, LLM translates the focal function based on the functionality implemented by the function, the dependencies used, and the syntax differences.

\begin{figure*}[b]
    \centering
    \includegraphics[width=0.8\linewidth]{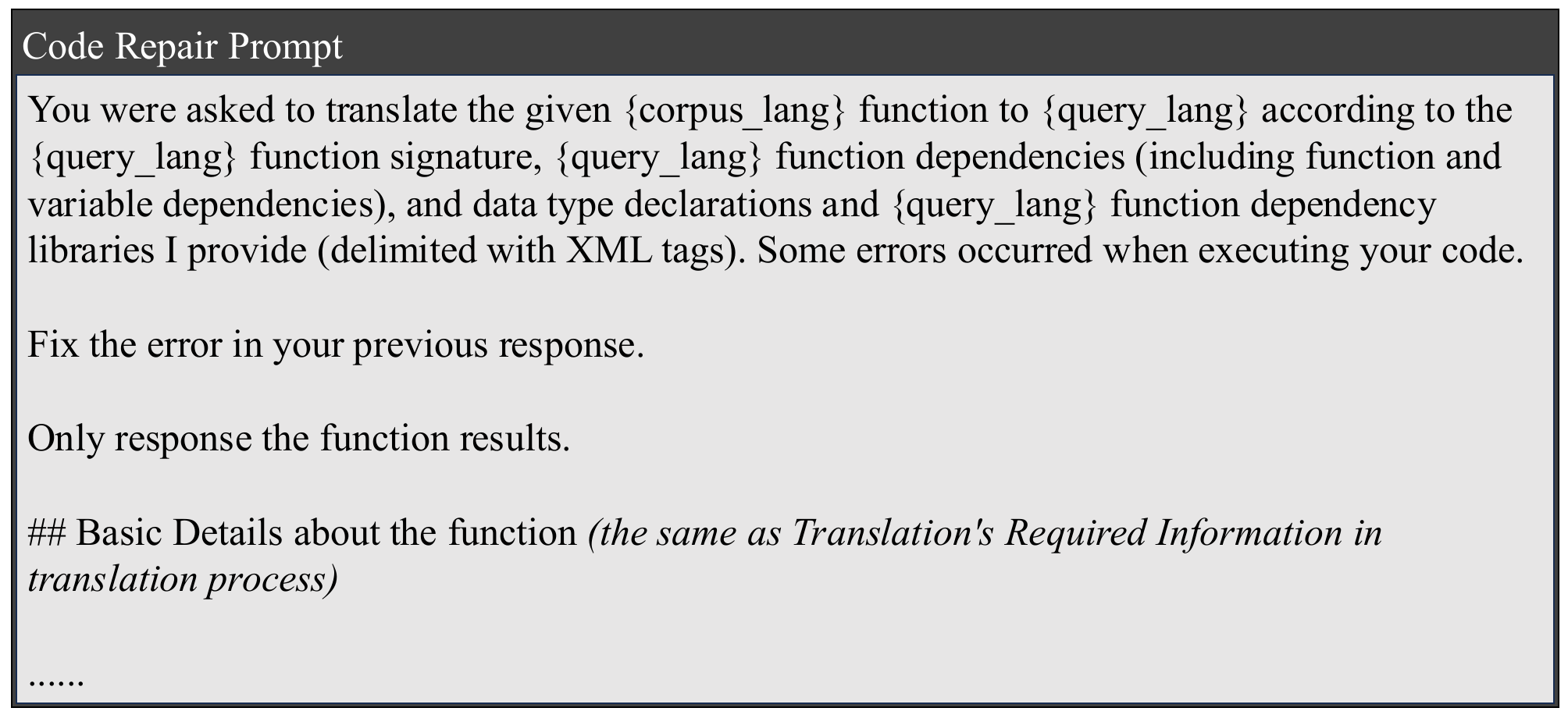}
    \vspace{-10pt}
    \caption{Prompt For Code Repair}
    \label{fig:prompt_code_repair}
    \vspace{-13pt}
\end{figure*}

\parabf{LLM-based code repair.} The code translated by LLMs may contain simple syntax errors (e.g., performing mutable operations on immutable variables). Previous work~\cite{deligiannis2024rustassistant} has shown that LLMs can repair certain errors in code using compiler error messages. After translation, \ourapproach provides the LLM with specific instructions to leverage error message and iteratively refine translation results that fail testing due to compilation or functional errors. While this process can be conducted for multiple iterations, our implementation limits it to one iteration for cost-effectiveness. This approach further improves translation accuracy and assesses the percentage of erroneous translations that can be successfully repaired. The prompt used for code repair is shown as Fig.~\ref{fig:prompt_code_repair}

\section{Experiment Setup}
We evaluate the effectiveness and usefulness of \ourapproach by answering the following five research questions:

\begin{itemize}[left=0em]
    \item \textbf{RQ1 (Performance of \ourapproach)}: How does \ourapproach perform compared to other LLM-based methods?

    \item \textbf{RQ2 (Ablation Study)}: How does each component of \ourapproach contributed to the final performance?

    \item \textbf{RQ3 (Impact of Retrieval Strategy)}: What is the impact of different retrieval strategies in Translation Knowledge Retrieval phase of \ourapproach?
    
    \item \textbf{RQ4 (Impact of Self-evolution)}: What is the impact of the self-evolution process on \ourapproach?

    \item \textbf{RQ5 (Bad Case Study)}: Why does \ourapproach fail in some translation tasks?
    
\end{itemize}

\subsection{Benchmark}
\label{sec:benchmark}
We choose RustRepoTrans~\cite{ou2024repository} for the following reasons. RustRepoTrans is the first repository-level context code translation benchmark, constructed from GitHub projects that reflect the complexities of real-world software development. It consists of 375 function translation tasks at the repository level, where the source languages are C, Java, and Python, and the target language is Rust. Each task is accompanied by corresponding test cases, with an average test coverage exceeding 90\%. As shown in previous work~\cite{ou2024repository}, LLMs exhibit much poorer performance on RustRepoTrans than on other benchmarks due to the repository-level dependencies and context. This shortcoming of LLM in dealing with real translation scenarios is exactly what \ourapproach aims to solve.

\parabf{Studied Programming Languages. }We focus on repository-level context code translation for translating code from C, Java and Python to Rust, which are the same supported language pairs in RustRepoTrans. A more important reason is that Rust, as the target language, is much closer to the actual requirements of real-world code translation scenarios~\cite{memory-safety}. Furthermore, without language-specific adaptation, if \ourapproach performs well with a low-resource language such as Rust, it can be reasonably inferred that \ourapproach would also be applicable to other high-resource languages.

\subsection{Baselines}

We evaluate the performance of \ourapproach with three baselines covering both LLM-based and rule-based methods: LLM-based methods include Basic RAG and LLM-based with prompt engineering (PE), while the rule-based method is represented by C2Rust.

\parabf{Basic RAG} is adapted from~\cite{bhattarai2024enhancingcodetranslationlanguage}, using the exact same prompt format and the same 1-shot strategy as \ourapproach to provide the most similar translation function example. The source of the translation function example is the remaining projects in the dataset excluding the current project, simulating the design in~\cite{bhattarai2024enhancingcodetranslationlanguage} where translation examples are drawn from an external pre-existing knowledge base. 

\parabf{LLM-based with PE.} Specifically, prompt engineering (as shown in Figure~\ref{fig:prompt_example}) refers to adding explicit instructions like \textit{`Do not perform a simple one-to-one translation; instead, consider the functional consistency of the code'} and translation CoT in the prompt with LLM-friendly format: markdown to guide the model not to perform a direct one-to-one translation, thereby improving its ability to recognize differences.
The difference between LLM-based with prompt engineering and \ourapproach is that the knowledge base in LLM-based with prompt engineering is empty, representing the cold-start translation phase.

\parabf{C2Rust~\cite{c2rust}, } the predominant automated tool for converting C code to Rust, operates exclusively at the syntax level. It leverages the fact that Rust supports both unsafe constructs—which align closely with those in C—as well as features that enforce memory safety to translate each feature used by the C code into a possibly unsafe equivalent feature of Rust to resolve the syntactic discrepancies between C and Rust.

\begin{table}[tbp]
\caption{Studied LLMs}
\vspace{-8pt}
\centering
\begin{adjustbox}{width=0.7\linewidth}
        \begin{tabular}{@{}c|c|cccc@{}}
            \hline
            \textbf{Model Type} & \textbf{Model Name } & \textbf{Open-source} & \textbf{reasoning} & \textbf{Time} & \textbf{Size} \\ \hline
            \multirow{3}{*}{General LLM} 
            & \claude      & \ding{55}  & \ding{55} & 2024.6 & -  \\ 
            & \vdeepseek    & \checkmark & \ding{55} & 2025.3 & 671B  \\ 
            & \rdeepseek & \checkmark & \checkmark & 2025.1 & 671B \\ \hline
            \multirow{2}{*}{Code LLM} 
            & \qwenthreetwo     & \checkmark & \ding{55} & 2024.9 & 32B  \\ 
            & \qwenfourteen     & \checkmark & \ding{55} & 2024.9 & 14B  \\  \hline

        \end{tabular}
        \end{adjustbox}
\label{tab:studied_llms}
\vspace{-10pt}
\end{table}

\vspace{-3pt}
\subsection{Metrics}
In line with previous works~\cite{roziere2020unsupervised,ou2024repository,khan2024xcodeeval,yan2023codetransocean}, we adopt the following match-based metrics, and execution-based metrics to evaluate the effectiveness of different repository-level code translation techniques.

\parabf{Match-based Metrics. }The matching-based metrics aim to evaluate the quality of generated code through static code analysis. We employed CodeBLEU and AST Match Score for evaluation:
\begin{itemize}[left=0em]

\item \textbf{\textit{CodeBLEU}}: CodeBLEU~\cite{ren2020codebleu} is a metric used to evaluate the quality of generated code, especially in the context of code generation tasks. It pays attention to the keywords, leverages the tree structure and considers the semantic logic information as expressed in Equation~\ref{eq:codebleu}.

\vspace{-6pt}
\begin{equation}
\text{CodeBLEU} = \alpha \cdot \text{BLEU} + \beta \cdot \text{BLEU}_{\text{weight}} + \gamma \cdot \text{Match}_{\text{ast}} + \delta \cdot \text{Match}_{\text{df}}
\label{eq:codebleu}
\end{equation}

\item \textbf{\textit{\astmatchscore}}: The \astmatchscore is part of CodeBLEU, the reason we extracted it separately is to evaluate the ability of LLMs to recognize syntax differences. 
If the model can accurately identify syntactic differences and avoid translating check statements that are required in the source language but unnecessary in the target language (e.g. null pointer checks in C, which do not need to be explicitly performed in Rust), the generated code paths will be more similar to the ground truth, thus resulting in a higher \astmatchscore.

\end{itemize}

\parabf{Execution-based Metrics. }The execution-based metrics aim to assess the quality of generated code through the execution status of the code. We employed Compilation@k, Pass@k, DSR@k and Repairable Rate for evaluation:

\begin{itemize}[left=0em]

\item \textbf{\textit{Compilation@k}}: Compilation@k as expressed in Equation~\ref{eq:CA} is a crucial evaluation metric that measures the proportion of translated code snippets that compiles successfully without errors within $k$ rounds, directly reflecting the syntactic and structural correctness of the translated code. A higher CA indicates that the model generates syntactically valid code that adheres to the grammar and compilation rules of the target programming language. $C(i, k) = 1$ if the $i^{th}$ code sample compiles successfully within $k$ attempts; otherwise $C(i, k) = 0$. In this study, \textit{Compilation@k} metrics were calculated with $k=1$, allowing for one generation attempt.

\vspace{-5pt}
\begin{equation}
\text{Compilation@k} =\frac{1}{N} \sum_{i=1}^{N} C(i, k)
\label{eq:CA}
\end{equation}

\item \textbf{\textit{Pass@k}}: This widely-used metric~\cite{chen2021evaluating} calculates the percentage of tasks correctly solved based on $k$ generated code samples per task, as expressed in Equation~\ref{eq:passk}. A task is considered solved if at least one of the generated code samples passes all the corresponding test cases. Following recent research~\cite{pan2024lost}, the focus was on calculating the \textit{Pass@1} metric, where $k=1$. This approach reflects the model's ability to produce a correct translation on the first attempt.
\begin{equation}
\text{Pass@k} = \mathbb{E}_{\text{Problems}} \left[ 1 - \frac{\binom{n - c}{k}}{\binom{n}{k}} \right]
\label{eq:passk}
\end{equation}

\item \textbf{\textit{DSR@k}}: The DSR@k (Debugging Success Rate@k) metric, proposed in previous work~\cite{yan2023codetransocean}, evaluates whether the generated code successfully executes and produces the expected results (\ie passing all test cases) within $k$ rounds of debugging, as expressed in Equation~\ref{eq:dsrk}. $S(i, k) = 1$ if the $i^{th}$ code sample succeeds within $k$ attempts; otherwise $S(i, k) = 0$. In this study, \textit{DSR@k} metrics were calculated with $k=1$, allowing for one debugging attempt.
\begin{equation}
\text{DSR@k} =\frac{1}{N} \sum_{i=1}^{N} S(i, k)
\label{eq:dsrk}
\end{equation}

\item \textbf{\textit{Repairable Ratio}}: Repairable Ratio (RR), as expressed in Equation~\ref{eq:RR}, is an evaluation metric for code translation tasks that measures the proportion of incorrect translations that can be automatically repaired using an LLM. 
The higher the value of this metric, the fewer interventions are needed to fix the code generated by the model, while initially incorrect, indicating higher quality of the generated code. 
\begin{equation}
\text{RR} =\frac{\text{DSR@1 - Pass@1}}{\text{1 - Pass@1}}
\label{eq:RR}
\end{equation}

\end{itemize}

\subsection{Implementations}
We construct the 100\% state of the knowledge base, where all functions except the target function have been successfully translated and can serve as sources for dependency usage examples and translation function pair examples, along with all currently available GitHub open-source repositories to build target-language code samples. This setup, referred to as \ourapproach in this paper, is used to evaluate the theoretical best performance that an LLM enhanced with self-evolving RAG can achieve in code translation tasks.
As shown in Table~\ref{tab:studied_llms}, the models we select as the foundational models for \ourapproach include general LLM and code LLM, open-source and closed-source models, reasoning models and non-reasoning models, as well as models with varying parameter sizes within the same series. This comprehensive selection demonstrates that \ourapproach is applicable to various existing types of large models.
However, due to resource constraints, after validating the generalizability of \ourapproach in RQ1, we selected the best-performing model, \claude, as the representative model for further evaluation and analysis in the subsequent RQs.
For the knowledge retrieval process, we utilize Elasticsearch~\cite{elasticsearch} as our search engine, which based on the Lucene library using BM25 as the default score function.

\section{Result}
\label{sec:result}

\subsection{RQ1: Performance of \ourapproach}
\label{sec:rq1}

\begin{table}[tbp]
\caption{Comparison of \ourapproach and LLM-based Baselines}
\centering
\begin{adjustbox}{width=1\linewidth}
        \begin{tabular}{@{}c|c|cccc|cc@{}}
            \hline
            \textbf{Model} & \textbf{Tech. } & \textbf{Compilation@1} & \textbf{Pass@1} & \textbf{DSR@1} & \textbf{RR} & \textbf{CodeBLEU} & \textbf{\astmatchscore} \\ \hline
            \multirow{3}{*}{\claude} 
            & Basic RAG      & 55.2\% & 48.3\%      & 69.9\% & 41.8\%        & 0.595   &  0.383  \\ 
            & PE      & 57.9\% & 49.1\%      & 69.6\% & 40.3\%        & 0.604   &  0.401  \\ 
            & \textbf{\ourapproach}& \textbf{75.7\%}& \textbf{67.7\%}& \textbf{82.1\%}  & \textbf{44.6\%} & \textbf{0.733}&  \textbf{0.562} \\ \hline
            \multirow{3}{*}{\rdeepseek}  
            & Basic RAG      & 57.3\% & 46.4\%      & 64.1\% & 33.0\%        &  0.588  & 0.375  \\ 
            & PE      & 61.1\% & 52.3\%      & 65.6\% & 27.9\%        & 0.605   &  0.400 \\
            & \textbf{\ourapproach}& \textbf{71.2\%}& \textbf{61.1\%}& \textbf{74.4\%}  & \textbf{34.2\%} & \textbf{0.725}&  \textbf{0.553} \\ \hline
            \multirow{3}{*}{\vdeepseek}  
            & Basic RAG      & 42.1\% & 35.2\%      & 48.5\% & 20.5\%        & 0.584   & 0.366  \\ 
            & PE      & 51.7\% & 43.7\%      & 51.5\% & 13.9\%        & 0.616   &  0.415  \\
            & \textbf{\ourapproach}&\textbf{74.4\%}& \textbf{63.2\%}& \textbf{71.5\%}  & \textbf{22.6\%} & \textbf{0.711}&  \textbf{0.529} \\ \hline
            \multirow{3}{*}{\qwenthreetwo}  
            & Basic RAG     & 31.2\% & 27.2\%      & 37.3\% & 13.9\%        &  0.576  &  0.352 \\ 
            & PE      & 36.3\% & 31.5\%      & 37.1 \% & 8.2\%        & 0.585   &  0.358  \\
            & \textbf{\ourapproach}& \textbf{53.6\%}& \textbf{47.5\%}& \textbf{56.8\%}  & \textbf{17.7\%} & \textbf{0.696}&  \textbf{0.503} \\ \hline
            \multirow{3}{*}{\qwenfourteen} 
            & Basic RAG      & 22.9\% & 19.2\%      & 33.3\% & \textbf{17.5\%}        & 0.515   & 0.278   \\ 
            & PE     & 25.6\% & 20.5\%      & 29.6\% & 11.4\%        & 0.465   &  0.233  \\ 
            & \textbf{\ourapproach}& \textbf{53.3\%}& \textbf{45.3\%}& \textbf{54.7\%}  & 17.2\% & \textbf{0.684}&  \textbf{0.493} \\ \hline
            
        \end{tabular}
        \end{adjustbox}
\label{tab:rq1_result}
\vspace{-10pt}
\end{table}

\parabf{Comparison with LLM-based baselines. }
As shown in Table \ref{tab:rq1_result}, \ourapproach achieved better results than both LLM-based baselines among all execution-based metrics for all studied LLMs. 
The best-performing model was \claude, with Compilation@1 and Pass@1 reaching 75.7\% and 67.7\%, respectively, substantially surpassing the best baseline PE by 17.8\% and 18.6\%. 
Furthermore, \ourapproach with \claude~achieved 82.1\% in DSR@1 and 44.6\% in RR, surpassing the baseline Basic RAG by 12.2\% and 2.8\%, respectively. This indicates that the code generated by \ourapproach is of higher quality with higher repairability.
In terms of match-based metrics, \ourapproach also significantly outperforms the best baseline, achieving 0.733 of CodeBLEU and 0.562 of \astmatchscore, representing substantial improvements over the baseline scores of 0.604 and 0.401. This fully demonstrates that the code generated by \ourapproach is more similar to the ground truth and better captures the syntactic differences between programming languages.

The most significant improvements were observed on \qwenfourteen, with relative increases of 108.2\% in Compilation@1 and 121.0\% in Pass@1. Additionally, CodeBLEU and Match scores improved by 47.1\% and 111.59\%, respectively. 
It should also be noted that the highest relative improvements of all metrics were all observed on \qwenfourteen, the worst-performing model. This indicates that \ourapproach can effectively compensate for the base model’s deficiencies in code translation, enhancing its capability in complex code scenarios.

In particular, LLMs combined with the instructions designed in \ourapproach outperform Basic RAG, which is adapted from previous work, in multiple metrics such as Pass@1 and CodeBLEU on all models. This indicates that even the prompt design in \ourapproach alone can effectively enhance LLM performance in repository-level context function translation.

\parabf{Comparison with Rule-based baseline. }
Since C2Rust only supports translating code from C to Rust, we extracted 145 tasks from RustRepoTrans where the source language is C and the target language is Rust, and compared the performance of \ourapproach and C2Rust on these tasks. Additionally, since rule-based methods lack self-debugging capabilities, we disregarded the DSR@1 and RR metrics. As shown in Table~\ref{tab:rq1_result_rule_based}, \ourapproach also achieved better results than the rule-based baseline across all execution-based metrics. C2Rust's Compilation@1 and Pass@1 were only 0.7\% and 0.7\% (i.e., successfully translating only one task). This is because rule-based methods operate at the syntax level and rely on one-to-one code mapping, making them unable to handle scenarios involving dependency differences. Among the failures, misinterpretations due to Function Dependencies Difference and Data Type Dependencies Difference accounted for as high as 95.2\% and 77.9\%, respectively. We will provide a more detailed analysis in Section~\ref{sec:rq5}. 
In terms of match-based metrics, the rule-based baseline also performed poorly, with CodeBLEU and \astmatchscore only 0.277 and 0.034, respectively. This is because C2Rust performs syntax-level, one-to-one translation of the source language code, resulting in low similarity with the target language version of the code.

\finding{\ourapproach outperforms all baselines across all metrics among all models. Even the prompt design in \ourapproach alone achieves slightly better results than the LLM-based baseline adapted from previous work. Furthermore, the rule-based method fails to effectively handle incremental translation scenarios with repository-level context.}

\begin{table}[tbp]
\caption{Comparison of \ourapproach and a Rule-Based Baseline on C-to-Rust Translation Tasks}
\centering
\begin{adjustbox}{width=0.85\linewidth}
        \begin{tabular}{@{}c|c|cc|cc@{}}
            \hline
             \textbf{Tech. } & \textbf{Model} & \textbf{Compilation@1} & \textbf{Pass@1} & \textbf{CodeBLEU} & \textbf{\astmatchscore} \\ \hline
             C2Rust & - &  0.7\%& 0.7\% & 0.277&  0.034 \\ \hline
             \multirow{5}{*}{\textbf{\ourapproach}} 
                     & \claude &  73.8\%& 61.4\%& 0.683&  0.490 \\ 
                     & \rdeepseek &  76.6\%& 62.1\%& 0.686&  0.486 \\ 
                     & \vdeepseek & 75.9\%& 57.9\% & 0.672&  0.466 \\ 
                     & \qwenthreetwo &  46.2\%& 37.2\% & 0.650&  0.423 \\ 
                     & \qwenfourteen &  51.0\%& 39.3\% & 0.640&  0.424 \\ \hline

        \end{tabular}
        \end{adjustbox}
\label{tab:rq1_result_rule_based}
\vspace{-10pt}
\end{table}

\begin{table}[b]
\caption{Ablation Study. K$^{\mathsf{1}}$ Trans means with Dependency usage examples, K$^{\mathsf{2}}$ Trans means with Dependency usage examples and Target-language Code Samples}
\resizebox{0.85\linewidth}{!}{%
\begin{tabular}{c|cccc|cc}
\hline
\textbf{Tech.}  & \textbf{Compilation@1} & \textbf{Pass@1} & \textbf{DSR@1} & \textbf{RR} & \textbf{CodeBLEU} & \textbf{\astmatchscore} \\ \hline
PE  & 57.9\% & 49.1\%       & 69.6\% & 40.3\%        & 0.604    &  0.401        \\ \hline
K$^{\mathsf{1}}$ Trans  & 72.3\%       & 62.9\%          & 78.6\% & 42.3\% & 0.630   &  0.432               \\ \hline
K$^{\mathsf{2}}$ Trans  & 73.9\%     & 65.6\%       & 79.2\%    & 39.5\%    & 0.708 &    0.533           \\ \hline
\textbf{\ourapproach}& \textbf{75.7\%}& \textbf{67.7\%}& \textbf{82.1\%}  & \textbf{44.6\%} & \textbf{0.733}&  \textbf{0.562}           \\ \hline
\end{tabular}

}
\label{tab:rq2_result}
\end{table}

\subsection{RQ2: Ablation Study}
\label{sec:rq2}

\ourapproach introduces three types of knowledge to improve LLM's ability of code translation: 1) dependency usage examples, i.e. providing relevant statement to teach LLM to correctly identify and invocate dependencies, 2) target-language code samples, i.e. retrieving similar target language code to lead LLM in generating code that conforms to the syntax of the target language, 3) the successful translation function pair, i.e. previously successful translation knowledge, enhancing the LLM's ability to recognize syntax differences between languages.

We explored the impact of different types of knowledge in the knowledge base on \ourapproach with \claude~by progressively reducing the provided knowledge categories. As shown in Table~\ref{tab:rq2_result}, each type of knowledge in the knowledge base significantly contributes to \ourapproach’s effectiveness in handling repository-level context code translation. With the inclusion of more types of knowledge, nearly every metric exhibits a clear trend of linear improvement.
In particular, the comparison between LLM-based with PE and K$^{\mathsf{1}}$ demonstrates that dependency usage examples lead to a substantial increase in execution-based metrics, with a maximum relative improvement of 28.1\%. 
This is because dependency usage examples provide the LLM with the correct invocation paths and methods for dependencies, effectively reducing the likelihood of LLMs incorrectly invoking dependencies, resulting in minimized compilation errors in the generated code, leading to more executable code. More fine-grained interpretation of the role of dependency usage examples with concrete instances are provided in Section \ref{sec:rq5}.

On the other hand, from the comparison results of K$^{\mathsf{1}}$ Trans and K$^{\mathsf{2}}$ Trans, we can know that target-language code samples have a significant impact on match-based metrics, leading to a 12.4\% and 23.4\% relative improvement in CodeBLEU and \astmatchscore, respectively. This demonstrates that target-language code samples enhance the static quality of the LLM-generated code that better conforms to the syntax and style of the target language.

\finding{Each knowledge significantly contributes to \ourapproach’s effectiveness in handling repository-level context code translation, with dependency usage examples making the most notable contribution.}

\subsection{RQ3: Impact of Retrieval Strategy}

\label{sec:rq3}
\begin{table}[t]
\caption{Impact of Retrieval Strategy. The retrieval method is BM25 combined with UniXcoder.}
\resizebox{0.85\linewidth}{!}{%
\begin{tabular}{c|cccc|cc}
\hline
\textbf{Retrieval Strategy} &  \textbf{Compilation@1} & \textbf{Pass@1} & \textbf{DSR@1} & \textbf{RR} & \textbf{CodeBLEU} & \textbf{\astmatchscore} \\ \hline
Top1         & 64.8\%       & 57.1\%         & \textbf{74.7\%}   & \textbf{41.0\%}   & 0.697               &   0.531             \\ \hline
Top2        & \textbf{68.0\%}                     & \textbf{57.6\%}       & 72.8\% &  35.8\%    & \textbf{0.706}         &    \textbf{0.536}            \\ \hline
Top3        & 67.2\%                     & \textbf{57.6\%}       &  73.6\% &  37.7\%   & 0.704         &    0.533             \\ \hline
\end{tabular}
}
\label{tab:rq3_result}
\end{table}

\begin{table}[]
\caption{Translation performance under different stages of the knowledge base: 0\%, 50\%, and 100\%. 100\% corresponds to \ourapproach.}
\resizebox{0.85\linewidth}{!}{%
\begin{tabular}{c|cccc|cc}
\hline
\textbf{Tech.}  & \textbf{Compilation@1} & \textbf{Pass@1} & \textbf{DSR@1} & \textbf{RR} & \textbf{CodeBLEU} & \textbf{\astmatchscore} \\ \hline
0\% knowledge base   & 57.9\% & 49.1\%       & 69.6\%    & 40.3\%      & 0.604                                      &  0.401  \\ \hline
50\% knowledge base   & 69.6\%       & 61.6\%         & 76.3\%   & 38.3\%   & 0.695                                &  0.507               \\ \hline
\textbf{100\% knowledge base}  & \textbf{75.7\%} & \textbf{67.7\%}       & \textbf{82.1\%}        & \textbf{44.6\%}  & \textbf{0.733} &    \textbf{0.562}           \\ \hline                                

\end{tabular}

}
\label{tab:rq4_result}
\end{table}

The scale of knowledge provided to the LLM influences its performance to some extent: too little knowledge may fail to cover critical information, preventing the LLM from generating correct answers, while too much knowledge may introduce noise that misleads the model, resulting in degraded performance. Therefore, we explore the optimal number of target-language code samples to provide in the repository-context code translation scenario.

As shown in Table~\ref{tab:rq3_result}, increasing the number of target-language code samples from one to two leads to improvements across Compilation@1, Pass@1, CodeBLEU, and \astmatchscore, with Compilation@1 increasing notably from 64.8\% to 68.0\%. This indicates that adding more target-language code samples initially has a positive effect on the LLM. However, when the number is increased to three, the metrics no longer show significant improvements, and some, such as Compilation@1, even decrease, suggesting that excessive target-language code samples can introduce noise.
Furthermore, as shown in RQ2 (Section~\ref{sec:rq2}), target-language code samples primarily impact match-based metrics like CodeBLEU and Match Score. Therefore, \ourapproach ultimately sets the number of target-language code samples to two, as achieving the best performance on Compilation@1, Pass@1, CodeBLEU, and \astmatchscore.

\finding{The optimal number of target-language code samples to provide in the repository-context code translation scenario is two.}

\subsection{RQ4: Impact of Self-evolution}

To evaluate the impact of the self-evolution process on \ourapproach, we design three knowledge base stages—0\%, 50\%, and 100\%—to simulate different stages of self-evolution. Here, 0\% corresponds to the baseline (LLM-based with Prompt Engineering), and 100\% corresponds to \ourapproach.
To comprehensively assess the practicality of the self-evolution process in different translation scenarios and to account for the randomness in the translation order of the target functions, we construct 50\% knowledge base as follows:
Dependency usage examples: For each target function, we randomly retain 50\% of the dependency usage examples available at the 100\% stage. The remaining 50\% dependencies have their dependency usage examples set to empty, representing no reference is available in current project’s context.
Target-language Code Samples: We randomly select 50\% projects from the collected projects and extract functions to construct the samples.
Successful translation function pairs: For each target function, we randomly select 50\% of the successful translation function pairs available at the 100\% stage. These selected pairs are then used as candidates for the top-1 successful translation function pair retrieval.

As shown in Table~\ref{tab:rq4_result}, when the knowledge base expands from 0\% to 50\%, execution-based metrics show significant improvements: Compilation@1, Pass@1, and DSR@1 increase by 11.7\%, 12.5\%, and 6.7\%, respectively, despite a slight decrease in Repairable Ratio. Similarly, match-based metrics exhibit substantial growth, with CodeBLEU increasing from 0.604 to 0.695 and \astmatchscore from 0.401 to 0.507.
When the knowledge base further expands from 50\% to 100\%, translation performance continues to improve, with relative improvement in various metrics ranging between 5.5\% and 16.4\%. As previous analyzed, the self-evolution process significantly enhances translation performance in both stages (0\% to 50\% and 50\% to 100\%). However, the improvements from 50\% to 100\% are less pronounced compared to the initial stage (0\% to 50\%). This suggests that after reaching 50\%, the remaining challenges may not be fully addressed through knowledge-driven enhancements alone, which we will explore in detail in RQ5 (Section \ref{sec:rq5}).

\finding{As the self-evolution process progresses, the knowledge base continuously enhances the LLM’s performance across various aspects of the repository-level code translation.}

\subsection{RQ5: \colortext{Case Study}}
\label{sec:rq5}

To fully understand the advantages and limitations of \ourapproach, we further manually analyze the successful examples of \ourapproach that the baselines fail, and the failure examples of \ourapproach.

\begin{table}[tp]
\vspace{-10pt}
\caption{The Number of Top3 Error Causes of Dependencies Misuse of \ourapproach and LLM-based Baseline}
\centering
\begin{adjustbox}{width=0.85\linewidth}
        \begin{tabular}{@{}c|c|ccc}
            \hline
            \multirow{2}{*}{\textbf{Model}}  & \multirow{2}{*}{\textbf{Tech.}} & \multicolumn{3}{c}{\textbf{Error type}}                   \\ \cline{3-5} 
            & & \begin{tabular}[c]{@{}l@{}}\textbf{Function Differences}\\\textbf{Misinterpretation ↓}\end{tabular}  & \begin{tabular}[c]{@{}l@{}}\textbf{Data Type Differences}\\ \textbf{Misinterpretation ↓}\end{tabular} & \begin{tabular}[c]{@{}l@{}}\textbf{Variable Differences}\\\textbf{Misinterpretation ↓}\end{tabular} \\ \hline

            \multirow{3}{*}{\claude} 
            & Basic RAG             & 368         & 2    & 131   \\ 
            & PE                    & \textbf{191}         & \textbf{0}    & 58   \\ 
            & \textbf{\ourapproach} & 247         & \textbf{0}    & \textbf{40}   \\ \hline
            \multirow{3}{*}{\rdeepseek}  
            & Basic RAG             & 228         & 1      & 48 \\ 
            & PE                    & \textbf{172}         & \textbf{0}      & 32\\
            & \textbf{\ourapproach} & 256         & 2      & \textbf{24}   \\ \hline
            \multirow{3}{*}{\vdeepseek}  
            & Basic RAG             & 420         & 1      & 188 \\ 
            & PE                    & 368         & \textbf{0}      & 100  \\
            & \textbf{\ourapproach} & \textbf{338}         & \textbf{0}      & \textbf{62}    \\ \hline
            \multirow{3}{*}{\qwenthreetwo}  
            & Basic RAG             & 679         & 11     & 220  \\ 
            & PE                    & 677         & 7      & 162  \\
            & \textbf{\ourapproach} & \textbf{447}         & \textbf{2}      & \textbf{83}   \\ \hline
            \multirow{3}{*}{\qwenfourteen} 
            & Basic RAG             & 1,288        & 11     & 219   \\ 
            & PE                    & 1,400        & 9      & 137 \\ 
            & \textbf{\ourapproach} & \textbf{524}         & \textbf{0}      & \textbf{64}   \\ \hline

            \multirow{3}{*}{Total} 
            & Basic RAG             & 2,983       & 25      & 806   \\ 
            & PE                    & 2,809       & 17      & 489 \\ 
            & \textbf{\ourapproach} & \textbf{1,812}       & \textbf{4}       & \textbf{273}   \\ \hline
            
        \end{tabular}
        \end{adjustbox}
\label{tab:rq5_case_study_good_case_compare_with_llm-based_baseline}
\vspace{-10pt}
\end{table}

\subsubsection{Analysis of Successful Examples}
Since LLM translation failures on RustRepoTrans are attributed to Dependencies Misuse errors in as high as 77.9\% of cases~\cite{ou2024repository}, we statistically analyzed the top three categories of Dependencies Misuse errors in the translation results of \ourapproach and all baselines:  
Function Differences Misinterpretation, Data Type Differences Misinterpretation, and Variable Differences Misinterpretation, recording the specific count for each category.

In terms of LLM-based baselines, as shown in Table~\ref{tab:rq5_case_study_good_case_compare_with_llm-based_baseline}, compared to all LLM-based baselines, \ourapproach effectively reduced the number of errors almost across all categories for all studied LLMs. Specifically, compared to the best-performing baseline PE, \ourapproach demonstrated a reduction of 35.5\% to 76.5\% across all error categories. This indicates that \ourapproach can significantly enhance the model's ability to accurately identify and handle non one-to-one mapping dependency differences in incremental translation scenarios.

Additionally, compared to PE, Basic RAG showed similar or even higher error counts across these three dependency error categories for all studied LLMs. This suggests that even the prompt design in \ourapproach (i.e., the PE baseline) alone slightly improves the model's capability to handle repository-level context compared to the Basic RAG baseline, which is consistent with the conclusions in Section~\ref{sec:rq1}.

\begin{figure*}[b]
    \vspace{-15pt}
    \centering
    \includegraphics[width=1\linewidth]{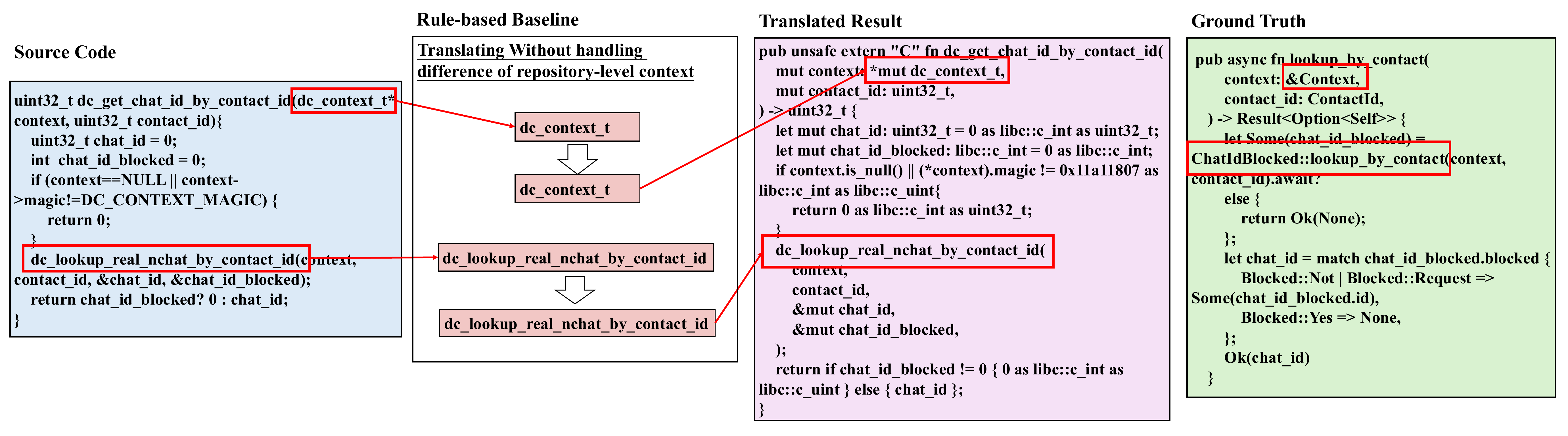}
    \caption{Failure Example From Rule-based baseline}
    \label{fig:bad_case_example_from_Rule-based_baseline}
    \vspace{-10pt}
\end{figure*}

In terms of the Rule-based baseline, as shown in 
Table~\ref{tab:rq5_case_study_good_case_compare_with_rule-based_baseline}, C2Rust exhibited 3,307 Function Difference Misinterpretation errors across 138 out of 145 tasks, and 637 Data Type Differences Misinterpretation errors across 113 tasks, affecting 95.2\% and 77.9\% of the tasks, respectively. The reason is that rule-based methods operate at the syntax level and rely on one-to-one code mapping, making them unable to handle scenarios involving dependency differences. Specifically, as shown in Fig.~\ref{fig:bad_case_example_from_Rule-based_baseline}, the Rule-based baseline performs a direct one-to-one translation of dependency call statements without any adaptation, failing to handle differences in repository-level context.
However, C2Rust demonstrated the best performance in Variable Differences Misinterpretation. This is because C2Rust translates non-local variables by statically parsing the specific values of variables in the source language and directly using these values instead of variable names in the target language. While this approach effectively reduces compilation errors potentially caused by variable differences, it significantly compromises code readability and does not actually address repository-level differences between the source and target language versions.

\begin{table}
\caption{The Number of Top3 Error Causes of Dependencies Misuse of \ourapproach and Rule-based Baseline}
\centering
\begin{adjustbox}{width=0.85\linewidth}
        \begin{tabular}{@{}c|c|ccc}
            \hline
            \multirow{2}{*}{\textbf{Tech.}}  & \multirow{2}{*}{\textbf{Model}} & \multicolumn{3}{c}{\textbf{Error type}}                   \\ \cline{3-5} 
            & & \begin{tabular}[c]{@{}l@{}}\textbf{Function Differences}\\\textbf{Misinterpretation ↓}\end{tabular}  & \begin{tabular}[c]{@{}l@{}}\textbf{Data Type Differences}\\ \textbf{Misinterpretation ↓}\end{tabular} & \begin{tabular}[c]{@{}l@{}}\textbf{Variable Differences}\\\textbf{Misinterpretation ↓}\end{tabular} \\ \hline

            {C2Rust} & - &  3,307 & 637 & \textbf{26} \\ \hline
            \multirow{5}{*}{\textbf{\ourapproach}} 
                     & \claude &  26 & \textbf{0}&  48 \\ 
                     & \rdeepseek &  \textbf{23}& \textbf{0}&  31 \\ 
                     & \vdeepseek & 35 & \textbf{0} &  34 \\ 
                     & \qwenthreetwo &  112& \textbf{0} &  145 \\ 
                     & \qwenfourteen &  129 & \textbf{0} &   127 \\ \hline
        \end{tabular}
        \end{adjustbox}
\label{tab:rq5_case_study_good_case_compare_with_rule-based_baseline}
\vspace{-10pt}
\end{table}

\subsubsection{Analysis of Failure Examples} 
In particular, we separately analyze the proportions of compilation errors and functional implementation errors in bad cases both before and after self-debugging. Furthermore, we classify compilation errors into finer-grained categories based on the taxonomy defined in RustRepoTrans~\cite{ou2024repository}. This analysis helps us identify which errors persist despite the triple-knowledge enhancement, as well as which errors can be easily fixed through self-debugging and which cannot. By doing so, we can pinpoint areas where future work should focus on improving LLM capabilities to further enhance performance in repository-level context code translation.

\begin{table}[t]
\caption{The Number of Each Error Causes of \ourapproach Before Self-debugging and After Self-debugging. One bad case can include more than one error}
\resizebox{0.85\linewidth}{!}{%
\begin{tabular}{c|c|cc}
\hline
\multirow{2}{*}{Error Type}                                                   & \multirow{2}{*}{Error Causes}                       & \multicolumn{2}{c}{Number}                   \\ \cline{3-4} 
                                                                              &                                                     & \begin{tabular}[c]{@{}c@{}}Before \\ Self-debugging\end{tabular} & \begin{tabular}[c]{@{}c@{}}After \\ Self-debugging\end{tabular} \\ \hline
\multirow{7}{*}{\begin{tabular}[c]{@{}c@{}}Compilation \\ Error\end{tabular}} & Data type Misinterpretation                         & 89                    & 54                   \\ \cline{2-4} 
                                                                              & Syntactic Differences Misinterpretation             & 0                     & 2                    \\ \cline{2-4} 
                                                                              & Function Differences Misinterpretation              & 309                   & 28                   \\ \cline{2-4} 
                                                                              & Variable Differences Misinterpretation              & 105                   & 16                   \\ \cline{2-4} 
                                                                              & Data Type Differences Misinterpretation             & 17                    & 9                    \\ \cline{2-4} 
                                                                              & Dependency Resolution Differences Misinterpretation & 0                     & 6                    \\ \cline{2-4} 
                                                                              & Function Signature Inconsistencies                  & 1                     & 0                    \\ \hline 
\end{tabular}
}
\label{tab:rq5_result}
\vspace{-10pt}
\end{table}

\begin{figure*}[b]
    \vspace{-15pt}
    \centering
    \includegraphics[width=1\linewidth]{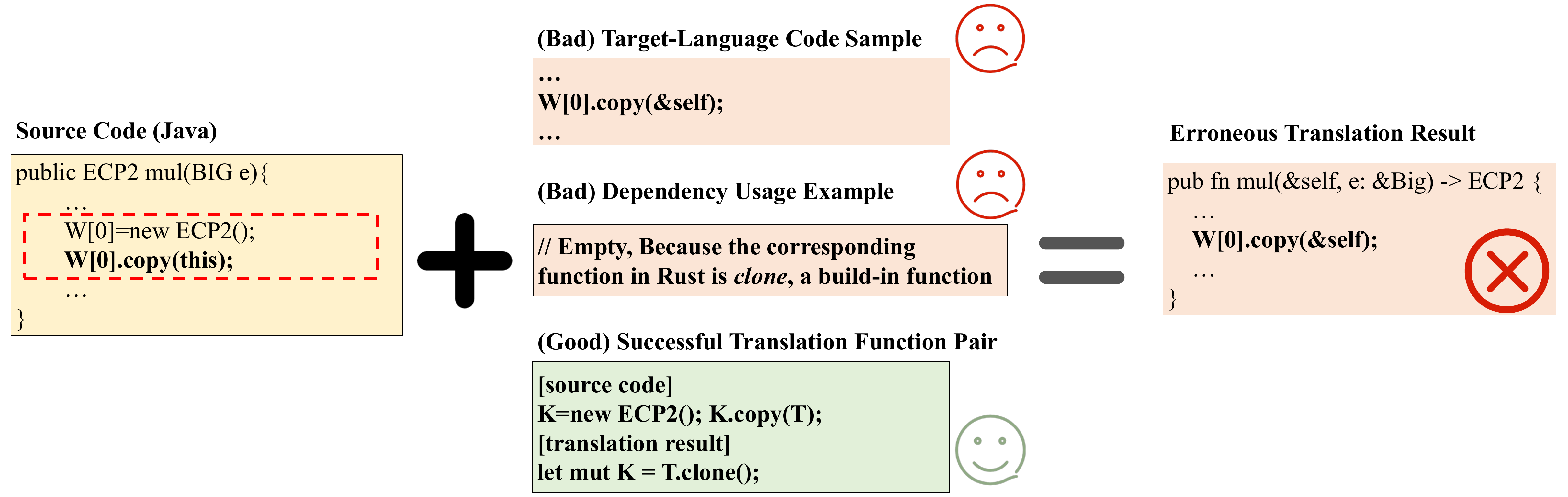}
    \caption{Failure Example From \ourapproach}
    \label{fig:bad_case_example}
    \vspace{-5pt}
\end{figure*}

\begin{figure}[tbp]
\vspace{-0.7em}
    \centering
    \begin{minipage}[t]{0.45\textwidth}
        \centering
        \includegraphics[width=1\textwidth]{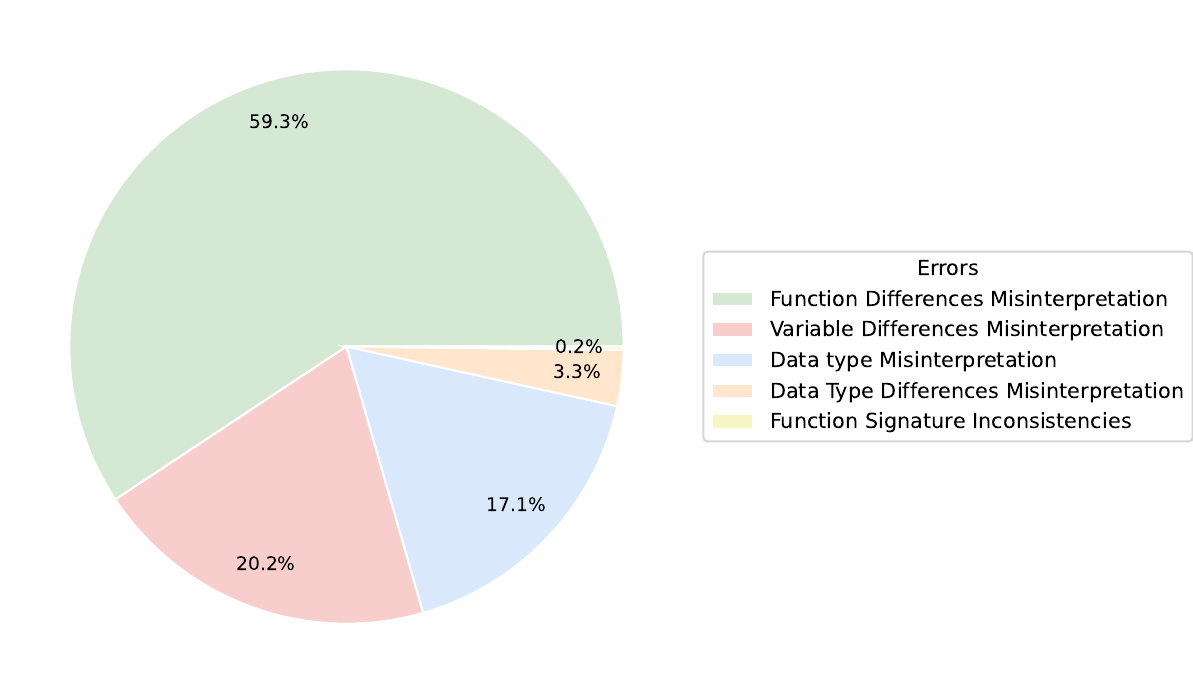}
        \caption{Distribution of Error Causes of \ourapproach before self-debugging}
        \label{fig:bug_type_propotion_before_debug}
        
    \end{minipage}
    \hspace{0.02\textwidth}
    \begin{minipage}[t]{0.45\textwidth}
        \centering
        \includegraphics[width=1\textwidth]{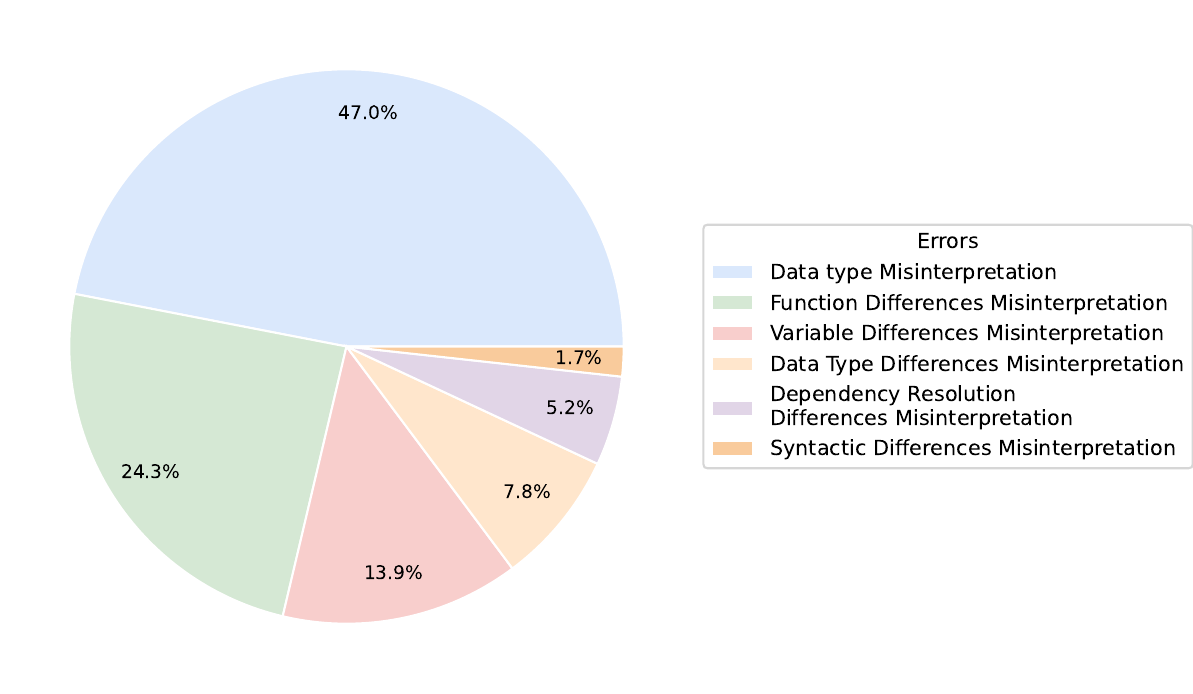}
        \caption{Distribution of Error Causes of \ourapproach after self-debugging}
        \label{fig:bug_type_propotion_after_debug}
    \end{minipage}
\vspace{-15pt}
\end{figure}

Before self-debugging, compilation errors account for 79.3\%, which is already significantly lower than the 94.8\% reported in~\cite{ou2024repository} evaluating Basic LLM on RustRepoTrans. This indicates that triple knowledge enhancement effectively reduces issues such as dependency misuse, syntax misinterpretation, and syntactical difference confusion, thereby improving compilation rates.
As shown in Figure~\ref{fig:bug_type_propotion_before_debug}, the highest proportion of compilation error types is Function Differences Misinterpretation, accounting for 59.3\%, with a total of 309 occurrences (detailed in Table~\ref{tab:rq5_result}). Through case analysis, we found that \ourapproach sometimes fails to provide LLM with fully corrective knowledge, particularly when the dependencies in the source project do not exist in the target project.
as shown in Figure~\ref{fig:bad_case_example}, for the statement under translation: \texttt{W[0].copy(this)}, \ourapproach fails to provide the correct knowledge in both Target-Language Code Sample and Dependency Usage Example. This is because the dependency \texttt{copy} in the source project does not have a direct counterpart in the target project but should instead be implemented by using the built-in function \texttt{clone} in the target language. Since \texttt{copy} is not implemented in the target project, \ourapproach cannot retrieve a valid dependency usage example for it.
Additionally, a higher textual similarity does not always guarantee a functionally equivalent match. As a result, in some cases, the retrieved code sample fails to provide the correct translation guidance. Therefore, even though Successful Translation Function Pair provides a correct example, the noise introduced by the other two types of knowledge misleads the LLM, ultimately leading to an incorrect translation.

After self-debugging, the proportion of compilation errors further decreased to 59.7\%. Notably, the most frequent compilation error before self-debugging—Function Differences Misinterpretation—dropped from 59.3\% to 24.3\%, with occurrences drastically reducing from 309 to just 28. This suggests that many of the existing shortcomings in \ourapproach can be effectively addressed by self-debugging.
It is important to note that Data Type Misinterpretation did not show a significant reduction after self-debugging. This is most likely because the target language (Rust) in the evaluated dataset is a low-resource language with stringent compile-time checks enforced by its ownership model and borrow checker. As a result, LLMs struggle with Rust-specific constraints in ways that cannot be easily mitigated by simply providing code samples. We will further explore and address this challenge in future work. 

\finding{\ourapproach fails to provide LLM with fully corrective knowledge in certain cases, such as when the dependencies in the source project do not exist in the target project. However, many of the existing shortcomings in \ourapproach can be effectively addressed by self-debugging.}

\section{Threats To Validity}
\parabf{Threats in benchmark.} There might be potential data leakage issue between the repository-level context code translation benchmark RustRepoTrans and sources of target-language code samples. 
However, we remove all projects with the same name as those involved in RustRepoTrans during projects collection, regardless of whether their creators are the same, considering the possibility of project forks. 
Furthermore, ablation study shows that even without target-language code samples, \ourapproach still significantly outperforms the baselines.

\parabf{Threats in generalization.} The only target language of the benchmark is Rust, which might limit the generalization of \ourapproach. 
However, our approach is not limited to Rust and can be extended to any other language in the future, as there is no language-specific adaptation in any phase of \ourapproach. 
Therefore, to apply \ourapproach to other programming languages, one only needs to collect relevant code for the target language and ensure the ability to analyze code in that language to extract target code sample. 
Extending the target language would be part of our future work.
\section{Related Work}

\label{sec:related}

\subsection{Code Translation}

Code translation is a vital tool in software development, enhancing interoperability and enabling code reuse. It is crucial for updating legacy systems and integrating components in different languages. Research shows that code translation improves productivity by reducing manual translation costs \cite{lu2021codexglue,puri2021codenet}, and supports the maintenance of multilingual projects.

Initially, rule-based systems were used in code translation, relying on manually defined syntax and semantic rules. While effective for simple translations, this approach struggled as projects became more complex. As research progressed, Statistical Machine Translation (SMT) introduced mapping common code patterns, improving accuracy in datasets like Java-C and CoffeeScript-JavaScript \cite{nguyen2013lexical,chen2018tree}. Subsequently, neural networks replaced rule-based systems, offering better flexibility. However, early models like Recurrent Neural Networks (RNNs) faced challenges with complex syntax \cite{dong2016language,yin2017syntactic,yin2018tranx}. 

Despite the progress made with LLMs, challenges such as incorrect translations and limited adaptability to coding styles persist \cite{zhu2022xlcost,rithy2022xtest}. This has led to a focus on evaluating translation quality and prompt engineering. Jiao et al. \cite{jiao2023evaluation} developed the G-TransEval benchmark, highlighting that models excel at simple tasks but struggle with complexity. Pan et al. \cite{pan2024lost} found that context-rich prompts improve reliability, while Yang et al. \cite{yang2024exploring} proposed strategies using test cases to boost accuracy. Macedo et al. \cite{macedo2024exploring} emphasized controlled prompts and post-processing for reliable benchmarking. Ou et al.~\cite{ou2024repository} proposed the first repository-level context code translation benchmark, narrowing the gap between existing evaluation datasets and real translation scenarios.
Our work focuses on the repository-level context translation task, which has not been fully considered in previous studies. By addressing this gap, we aim to enhance the practical adoption of LLMs in industrial applications, ensuring that they can generate more accurate and executable translations within real-world software development environments.

\subsection{Retrieval-Augmented Generation}

Traditional LLMs like GPT and BERT rely on their training knowledge to handle tasks, exhibiting strong text generation but facing challenges like outdated knowledge, lack of domain expertise, and "hallucinations" in specialized tasks. RAG addresses these by integrating external information sources into the model's process \cite{Khandelwal2019GeneralizationTM, Min2020AmbigQAAA}. Specifically, RAG adds a retriever to the generative model, retrieving relevant data from external sources to improve performance.

RAG shows great potential in code translation tasks. Traditional models like Seq2Seq, which rely on large code pairs for training, face limitations. For example, the Seq2Seq model by Acharjee et al. \cite{Acharjee2022SequencetoSequenceLB} was trained on the SPoC dataset containing pseudocode and source code pairs, which helps improve translation accuracy. However, such models often struggle with limited training data and emerging language features. In contrast, RAG overcomes these limitations by enhancing translation through dynamic retrieval of information from code repositories and documentation, which significantly improves both accuracy and quality.

However, RAG faces challenges. One is maintaining a retrieval database, as Lin et al. \cite{Lin2024DomainAA} suggest addressing domain-specificity and knowledge base heterogeneity through continuous pretraining and a unified knowledge format. Another challenge is ensuring the retriever quickly identifies relevant documents, as Sawarkar et al. \cite{Sawarkar2024BlendedRI} demonstrate by using semantic search and blended query strategies to improve retrieval performance.
Our work constructs a self-evolving strategy to continuously expand the translation knowledge base to consistently improve translation quality of LLMs.
\section{Conclusion}

In this work, we propose a novel LLM-based code translation technique \ourapproach, which leverages evolving triple knowledge augmentation to enhance LLM’s translation quality under repository context in real-world software development. 
The experiments show that \ourapproach substantially outperforms the baseline adapted from previous work, achieving relative improvements of up to 135.9\% on Pass@1 and 32.8\% on CodeBLEU among studied LLMs. Furthermore, the code generated by \ourapproach is of higher quality, as indicated by the higher DSR@1 and Repairable Ratio, which suggests a greater proportion of fixable code.
It is important to note that the results also demonstrate that each knowledge significantly contributes to \ourapproach’s effectiveness in handling repository-level context code translation, with dependency usage examples making the most notable contribution. 
Moreover, as the self-evolution process progresses, the knowledge base continuously enhances the LLM’s performance across various aspects of the repository-level code translation.

\bibliographystyle{ACM-Reference-Format}
\bibliography{ref}

\end{document}